\documentclass[11pt,twoside]{article}

\newif\ifpdf\ifx\pdfoutput\undefined\pdffalse\else\pdfoutput=1\pdftrue\fi

\usepackage{graphicx}
\usepackage{asp2006}
\usepackage{epsf}
\usepackage{psfig}
\usepackage{lscape}

\begin{document}
\title{Calibration and imaging challenges at low radio frequencies: An
  overview of the state of the art}
\author{S. Bhatnagar} \affil{National Radio Astronomy Observatory,
  Socorro, New Mexico, USA} 

\begin{abstract}
Many scientific deliverables of the next generation low frequency
radio telescopes require high dynamic range imaging.  Next generation
telescopes under construction indeed promise at least a ten-fold
increase in the sensitivity compared with existing telescopes.  The
projected achievable RMS noise in the images from these telescopes is
in the range of 1--10$\mu$Jy/beam corresponding to typical imaging
dynamic ranges of $10^{6-7}$.  High imaging dynamic range require
removal of systematic errors to high accuracy and for long integration
intervals.  In general, many source of errors are directionally
dependent and unless corrected for, will be a limiting factor for the
imaging dynamic range of these next generation telescopes.  This
requires development of new algorithms and software for calibration
and imaging which can correct for such direction and time dependent
errors.  In this paper, I discuss the resulting algorithmic and
computing challenges and the recent progress made towards addressing
these challenges.
\end{abstract}

\section{Introduction}

Aperture synthesis array telescopes combine signals from a number of
antenna pairs to sample the coherence function (visibility function)
in the radiation far field.  The angular resolution is inversely
proportional to the largest separation between the antennas (baseline)
compared to the wavelength of observation.  The sensitivity is
proportional to the total collecting area, square root of the
bandwidth of observation and total integration time and inversely
proportional to the effective system temperature.

Next generation telescopes radio telescopes, some under construction,
promise 10--100 times improvement in resolution and sensitivity.  For
a number of reasons ranging from engineering challenges and cost
considerations to sky background emission, it is hard to lower the
system temperature by a few order of magnitude to improve the
telescope sensitivity by similar order.  As a result, all next
generation telescopes use larger number of antenna elements to
increase the collecting area, wide-band receivers and long
integrations in time to achieve the higher sensitivity.  To mitigate
bandwidth smearing \citep{THOMPSON_AND_MORAN}, effects of narrow band
radio frequency interference (RFI) and for scientific reasons, the
observed band is split into a number of narrower frequency channels.
Snapshot data rate is proportional to the product of the square of the
number of antenna elements and number of frequency channels.  This
fact, combined with long integrations in time implies that the
projected sensitivity improvements of the next generation telescopes
will come at the cost of $10^{2-4}$ times increase in the data volume
over existing telescopes.

An underlying assumption in the sensitivity calculations is that the
random noise in the observations has no systematic component and that
for a given system temperature, the signal to noise ratio (SNR) is
proportional to $\sqrt{\Delta T~\Delta \nu}$ where $\Delta T$ is the
total integration in time and $\Delta \nu$ is the total bandwidth of
observation.  However, the observed data is inevitably corrupted by a
number of instrumental and ionospheric/atmospheric effects.
Furthermore, these effects are not the same across the field of view
(i.e., these effects are direction dependent(DD)).  This makes the
``noise'' in the observations non-random which does not necessarily
reduce with integration in time and/or frequency.  Furthermore, low
frequency sky is also brighter and more complex.  As a result, the
projected image plane RMS noise of 1--10$\mu$Jy/beam translates to an
imaging dynamic range requirement of $10^{6-7}$.  The imaging dynamic
range limit due to deconvolution errors for complex fields with
compact and extended emission is significantly higher than this.

The next generation post processing software therefore needs to
correct for direction dependent effects more accurately, over larger
parameter space (time, frequency and polarization) using 2--4 orders of
magnitude larger data volume as well as image complex sky emission
with high fidelity to achieve the scientific goals.  An obvious
conclusion is that we necessarily need significant research in the
area of post processing techniques for imaging and calibration and
develop algorithms which are more accurate, account for direction
dependent effects and can deal with large data volumes efficiently.
In this paper, I review the recent progress in the development of new
imaging and calibration algorithms relevant for high dynamic range
imaging at low radio frequencies ($< 2$ GHz).  A more complete
theoretical background can be found in the recent paper by
\cite{IMAGING_THEORY_IEEE}.

\section{The Measurement Equation}
\label{Sec:ME}
Using the theoretical formulation by \cite{HBS1}, full polarimetric
measurements from a single baseline can be described by the following
Measurement Equation
\begin{equation}
\label{ME}
\vec{V}^{Obs}_{ij}(\nu,t) = J_{ij}(\nu,t) W_{ij}(\nu,t) \int E_{ij}(\vec{s},\nu,t)
\vec{I}(\vec{s},\nu,t) e^{\iota \vec{b}_{ij} \cdot \vec{s}}d\vec{s}
\end{equation}
where $\vec{V}^{Obs}_{ij}$ is the observed visibility samples measured
by the pair of antennas designated by the subscript $i$ and $j$,
separated by the vector $\vec{b}_{ij}$ and weighted by the measurement
weights $W_{ij}$.  $J_{ij}$ is the complex {\it direction independent}
gain, $E_{ij}$ is the {\it direction dependent} gain as a function of
the direction $\vec{s}$, frequency $\nu$ and time $t$ and $\vec{I}$ is
the image vector.  The vectors $\vec{V}$ and $\vec{I}$ are full
polarization vectors in the data and image domain respectively.
$J_{ij}$ and $E_{ij}$ can be expressed as an outer product of two $2
\times 2$ antenna based Jones matrices as $J_{ij}=J_i(\nu,t) \otimes
J^{*}_j(\nu,t)$ and $E_{ij}=E_i(\vec{s},\nu, t) \otimes
E^{*}_j(\vec{s},\nu,t)$. $J_i$ and $E_i$ describe the full
polarization response of the individual antennas in the feed
polarization bases.  An appropriate unitary transform can be applied
to convert the above equation to Stokes bases (see \cite{HBS1,
IMAGING_THEORY_IEEE} for details).

For wide-band observations, the sky emission also changes as a
function of frequency and is potentially differently for different
directions.  Assuming time invariance, this frequency dependence can
be expressed as
\begin{equation}
\label{WBSky}
\vec{I}(\vec{s}, \nu) = \vec{I}^o(\vec{s},\nu_o)
\left(\frac{\nu}{\nu_o}\right)^{\alpha(\vec{s},\nu)}
\end{equation}
where $I^o$ is the image at the reference frequency $\nu_o$ and
$\alpha$ is spectral index which varies across the field of view and
the frequency band.  $I^o$, $J_i$, $E_i$ and $\alpha$ represent the
{\it explicit} unknowns in Equation~\ref{ME}.  The process of
calibration estimates $J_i$ and $E_i$ while the process of imaging
estimates $I^o$ and $\alpha(\vec{s},\nu)$.  Note that while the
observed visibilities can be corrected for the effects of $J_i$s by
dividing Equation~\ref{ME} by $J_{ij}$, the same is not true for DD
terms.  Correction for the effects of $E_{ij}$ can only be done as
part of the imaging process.  This makes solving and correcting for DD
errors more difficult and consequently conventional calibration
accounts for only direction independent corruptions.  This has been
sufficient for the existing telescopes.  This however must change to
achieve imaging dynamic ranges consistent with the thermal noise limit
of the next generation instruments \citep{POINTING_SELFCAL,
AWProjection_Memo,MSMFS_Memo}.

\section{Parametrization of the Measurement Equation}
\label{Sec:PME}
In order to make an image free of the effects of the corruptions,
$\vec{V}^{Obs}$ needs to be corrected for the effects of $J_i$s and
$E_i$s. Conventional techniques typically parametrize $J_{ij}$ as
three separate terms to represent time, frequency and polarization
dependencies.  These are assumed to be orthogonal and therefore solved
independently in the process of time, bandpass and polarization
calibration.  For narrow band observations (less than 10\% fractional
bandwidth), $\alpha$ is assumed to be small or zero and $I^o$ is
parametrized as a value per pixel, each pixel being treated as
independent degree of freedom (DoF).  $E_i$ is ignored in calibration
and during image deconvolution and if required, corrections for it are
made post-deconvolution (e.g. post-deconvolution correction for the
antenna response as a function of direction).

For the next generation telescopes however, more sophisticated
parametrization is required.  The DD terms need to be parametrized to
model the instrumental and ionospheric DD effects. Variations of
$\alpha$ across observing frequency band and across the field of
view (FoV) needs to be parametrized to model the spectral index of
the sources.  Sky is brighter and more complex at low frequencies and
most fields have sources with extended emission. $I^o$ therefore also
needs to be parametrized to better represent extended emission.

\begin{figure*}[ht!]
\hskip -0.25in
\hbox{
\includegraphics[width=7cm]{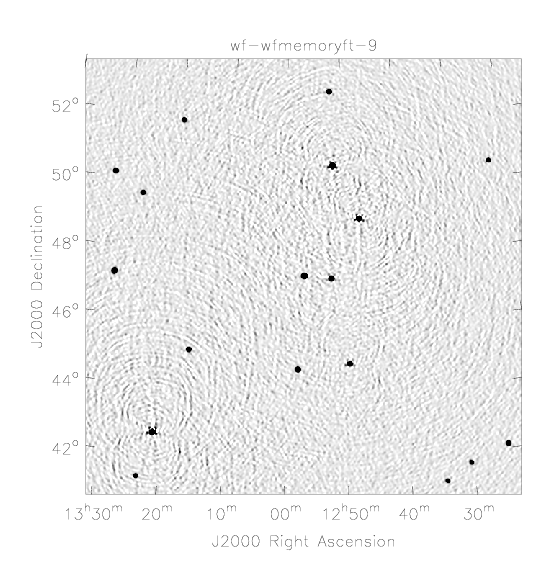}
\includegraphics[width=7cm]{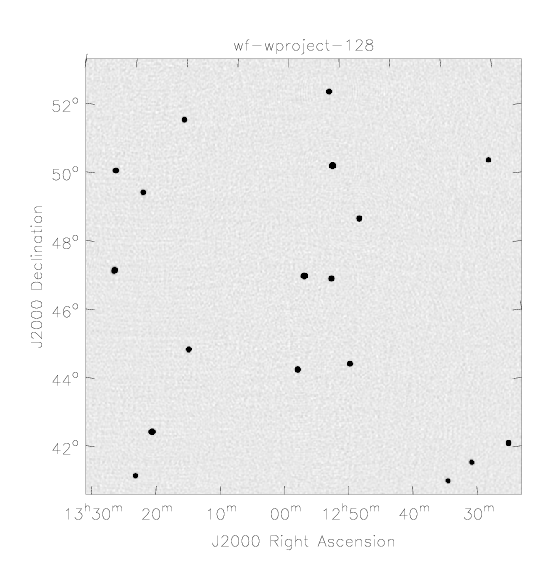}
}
\caption{\small Figure shows the performance of imaging algorithms to
  correct for the effects of the w-term.  Image on the left was made
  using the uv-faceting algorithm.  Image on the right was made using
  the w-projection algorithm.  Compact sources well away from the
  center of these images are undistorted.  The RMS noise in the two
  images is the same.  Residual errors are more systematic for the
  facet based algorithms.}
\label{Fig:WTerm}
\end{figure*}
\begin{figure*}[ht!]
\centering
\includegraphics[width=7cm]{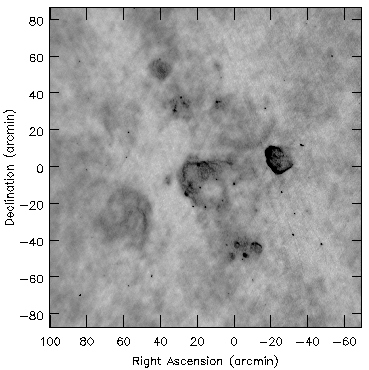}
\caption{\small Image showing wide-field P-band imaging in the
  Galactic plane using the W-Projection algorithm.  Extended emission
  as well as compact sources away from the phase center show no
  distortions due to the w-term.}
\end{figure*}
\section{The W-term}

The exponent in Equation~\ref{ME} can be expanded as
$\vec{b}_{ij}\cdot \vec{s} = u_{ij} l + v_{ij} m +
w_{ij}\left[\sqrt{1-l^2-m^2}-1\right]$ with the usual meaning for
the symbols \citep{THOMPSON_AND_MORAN}.  When the field of view is
large (as is typically true at low frequencies), the integral cannot
be reduced to a Fourier transform and use of the 2D FFT algorithm for
computing efficiency leads to significant distortions away from the
phase center.

Faceting algorithms using faceting in the uv-domain (as against
image-plane faceting) (see \cite{UVFACETING} for an expression for
faceting in the uv-domain) produce undistorted single-plane images of
the sky with an approximate space-invariant PSF.  This has several
practical advantages during deconvolution, particularly for the
deconvolution of extended emission. Algorithms using this approach
exist in the CASA package\footnote{CASA Home Page at {\tt
http://casa.nrao.edu}}.  Computing load is the same as that for
image-plane faceting algorithms \citep{TIM_N_RICK_1992,WFI_LECTURES2}.

\subsection{The W-Projection Algorithm}
\label{Sec:WProj}
Absorbing the third term in the expression for $\vec{b}_{ij}\cdot
\vec{s}$ into $E_{ij}$ in Equation~\ref{ME}, the observed visibilities can
be expressed as $\vec{V}^{Obs}_{ij}= \left[FT\left(E_{ij}\right)\right]
\star \vec{V}^o$ where $V^o$ are the visibilities corresponding
to the tangent plane, $FT$ represents the Fourier transform and
'$\star$' denotes the convolution operations respectively.  The
W-Projection algorithm \citep{W_Projection_IEEE} exploits this to
correct for the w-term in the gridding/de-gridding operation during
imaging.  While theoretically this algorithm can be shown to be faster
by up to 50 times, in practice it has been shown to be up to an order
of magnitude faster compared to faceting algorithms.  W-Projection
algorithm also produces a single-plane image, making it easier to
combine with other techniques for dealing with extended emission
across the field of view as well as correcting for other DD effects.

\section{Ionospheric corruptions}

Corruptions due to ionosphere is one of the limiting problems in high
sensitivity high resolution imaging at low frequencies.  Its effect is
that the phase across the antenna aperture is not constant and
potentially different for each antenna in the array (it is a direction
dependent effect - often referred to as ``non-isoplanatic ionosphere''
in the literature).

\subsection{Field base calibration}

The field-based calibration technique \citep{FieldBasedCalibration2}
measures the shift of compact sources throughout the FoV to estimate
local ionospheric phase gradients.  A polynomial fit to the estimated
phases is then used to apply corrections to the rest of the field.
For small baselines ($< 2-3$Km) this has been shown to improve the
imaging performances for some fields.  This technique however can be
computationally prohibitive for large FoV with complex emission and
does not extend to cases where the ionospheric refractive effects are
significant.

\subsection{Peeling}
\label{Sec:Peel}
This technique estimates a complex gain towards a number of sources
across the FoV for which good models are known apriori (either from
earlier imaging and calibration runs or from external sources).  The
solved gains are then used to remove the contribution of the sky
emission in the vicinity from the data as:
\begin{equation}
\vec{V}^{Corrected}_{ij} = \vec{V}^{Obs}_{ij} - \sum_k J_{ij}^P
\vec{V}_{ij}^{k^{Model}} 
\end{equation}
where the superscript $k$ denotes all sources in the region where the
peeling solutions $J_{ij}^P$ apply.  In the iterative form of Peeling,
the corrected visibilities are then used to apply this technique
iteratively to the strongest sources in the residual image.  This has
been shown to work well for simple fields (dominated by compact strong
sources) and for relatively small data volume.  Various variants of
this technique are being currently tested \citep{Peeling_LOFAR,
Peeling_MWA} to determine its numerical and computational performance
for complex fields with extended emission and with large data volumes.


\section{Effects of antenna Primary Beam}
\label{Sec:PB}
For narrow band observations, time dependent DD gains are dominantly
due to time varying antenna primary beams (PB).  While the antenna
forward gain is clearly direction dependent, its time variation is due
to a number of reasons (rotation of the rotationally asymmetric PBs
with Parallactic Angle for Az-El mount antennas, antenna pointing
errors, geometrical distortions of the antenna with elevation, etc.).
For wideband observations, the shape of the PB varies across the band
and sources well within the PB main-lobe at the lower frequency end of
the band may appear in the first sidelobe at the higher frequency end
(e.g. with a bandwidth ratio of 2:1).  For sources in the first
sidelobe of the antenna power pattern, the time varying gain due to
the rotation of the PBs will be even stronger
\citep{AWProjection_Memo,AWProjection}.  Furthermore, the time,
frequency and direction dependence of aperture array station power
pattern is expected to be worse compared to filled aperture antennas.
Algorithms to correct for PB effects are therefore crucial for the
scientific deliverables of the next generation instruments.

Image-plane based PB correction by direct evaluation of the integral
in Equation~\ref{ME} is possible.  To reduce the resulting prohibitive
run-time computing cost for realistic image complexity and data
volumes, a FFT based reverse transform has been used by
\cite{USON_N_COTTON2008}.  However this requires making assumptions
about the variability of either the sky emission or the antenna power
pattern.
\subsection{The A-Projection Algorithm}
\label{Sec:AProj}
\begin{figure*}[ht!]
\hbox{
\hskip 0.2in
\includegraphics[width=6cm]{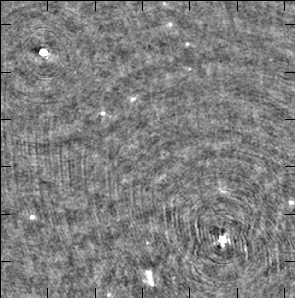}
\includegraphics[width=6cm]{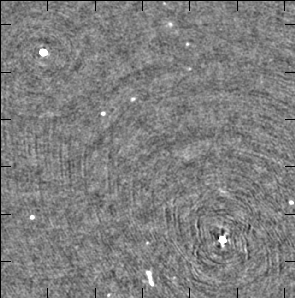}
}
\hbox{
\hskip 0.2in
\includegraphics[width=6cm]{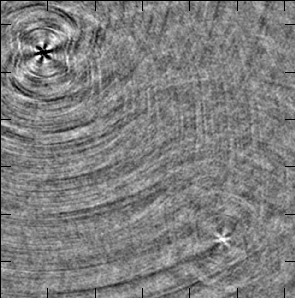}
\includegraphics[width=6cm]{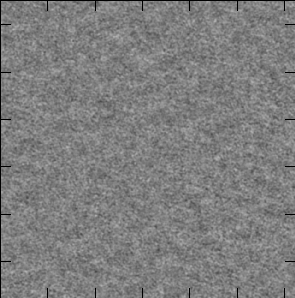}
}
\caption{\small Figure shows the results of the application of the
  A-Projection algorithm for VLA L-Band imaging. Top panel shows
  Stokes-I images made using conventional (left) and the A-Projection
  algorithm (right).  Bottom panel shows Stokes-V images.  Stokes-V
  imaging with the VLA suffers from strong and time varying
  instrumental effects which are completely corrected in the image
  in bottom right panel.}
\label{Fig:AProj}
\end{figure*}

$E_{ij}$ represents the effects of antenna primary beams in
Equation~\ref{ME}.  The A-Projection algorithm \citep{AWProjection} uses a
model for antenna aperture illumination and the approximate unitary
nature of the resulting operator to correct for effects of PB as part
of image deconvolution iterations.  This algorithm can naturally deal
with non-identical antenna PBs, is straight forward to integrate with
other advanced imaging and calibration algorithms and is
computationally efficient.  An estimate of the antenna aperture
illumination pattern is however required - which can be measured
(antenna holography) or modeled.  See Fig.~\ref{Fig:AProj} for an
example of application of this algorithm for imaging at L-Band using
the VLA.


\section{The Sky Model}

The sky model $\vec{I}$ in Equation~\ref{ME} is computed using image
deconvolution algorithms.  Most conventional algorithms treat each
pixel with significant emission in the image as an independent DoF
(e.g., the Clean \citep{Hogbom_Clean} and MEM \citep{MEM} algorithms
and their variants).  Such a parametrization of the sky is non-optimal
for extended emission \citep{Asp_Clean} and suffers from the problem of pixel
quantization errors \citep{Pixelation_Errors,
Pixelation_Errors_Cotton}.  For complex fields with strong emission,
both these problems limit the imaging dynamic range well above the
instrumental limit of the next generation telescopes.

\begin{figure*}[ht!]
\centering
\hskip 0.25in
\hbox{
\includegraphics[width=4cm,height=4.15cm]{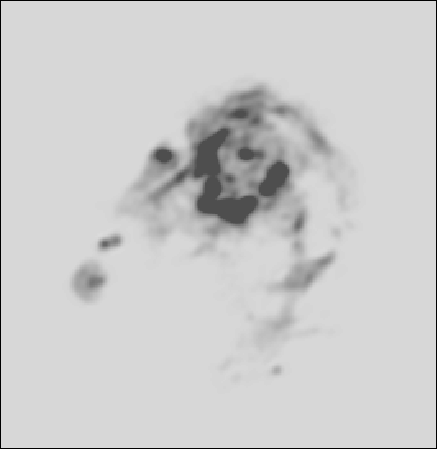}
\includegraphics[width=4cm]{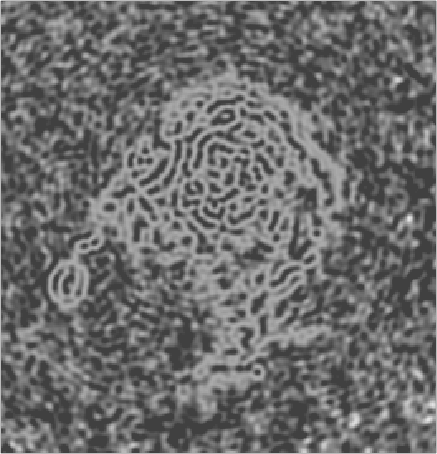}
\includegraphics[width=4cm]{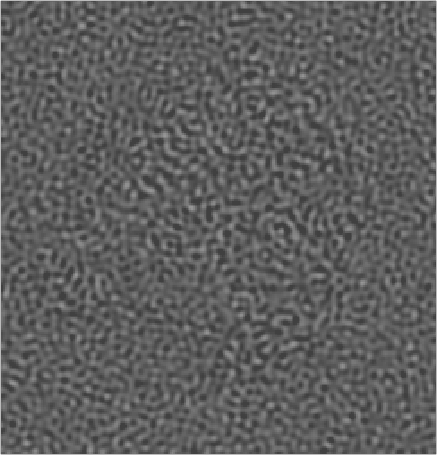}
}
\caption{\small Figure showing the performance of MS-Clean and
  Asp-Clean algorithms for the deconvolution of complex extended
  emission.  Image on the left is the model image used for the tests.
  Images in the middle and on the right show the residuals from
  MS-Clean and Asp-Clean algorithms respectively.  Spatial correlation
  scale is significantly reduced in both residuals compared to
  residuals from the Clean algorithm (not shown).  Residuals from
  Asp-Clean algorithm are more noise like compared to MS-Clean
  algorithm.}
\label{Fig:FieldBasedCal}
\end{figure*}

\subsection{Scale-sensitive modeling: The MS- and Asp-Clean algorithms}

The MS-Clean algorithm \citep{MSCLEAN} models the sky as a collection
of components with pre-computed set of scale sizes.  Depending upon
the user defined choice of scales, extend emission as well as compact
emission can be better modeled with a significantly smaller number of
components (DoF).  Memory requirements and computing load is higher
compared to conventional algorithms and the coupling between fixed set
of user defined scales is ignored (i.e., it effectively ignores the
fact that the parameter space is non-orthogonal).

The Asp-Clean algorithm \citep{Asp_Clean} adaptively determines the
local scale as well as position of the components in the image.  This
effectively mitigates the problem of pixel quantization and accounts
for coupling between various scales (i.e., recognize the fact that
parameter space is non-orthogonal).  Although for the same number of
components the computing requirements are 2--3 times higher compared
to MS-Clean, the number of components required is significantly
smaller for same image complexity.

\subsection{Wide-band modeling: The MS-MFS algorithm}

Ignoring the frequency dependence of the sky in wide-band observation
can limit the imaging dynamic range to $\sim 1:10^4$
\citep{MSMFS_Memo}.  Hence, apart from scale-sensitive modeling of the
sky, modeling the frequency dependence of the sky
(Equation~\ref{WBSky}) during imaging is required for wide-band
observations. The MS-MFS algorithm \citep{RAU_THESIS} uses the
MS-Clean approach to model extended emission and models the frequency
dependence using a Taylor expansion of Equation~\ref{WBSky} about the
reference frequency and solving for the coefficients of the series.
The A-Projection algorithm is used to correct for the PB frequency
dependence in combination with MS-MFS to make Stokes-I, spectral index
and spectral index variation images of the sky \citep{RAU_THESIS}.
The combined algorithm is being currently tested using wide-band
observations of fields with strong extended emission.


\section{Solvers for direction dependent effects}

For high resolution high dynamic range imaging, it is virtually
impossible to measure the DD terms in Equation~\ref{ME} to the
required accuracy prior to imaging.  Algorithms to model and solve for
the DD effects are therefore required.

Peeling based algorithms attempt to solve for DD effects by allocating
few DoF per direction of interest (DoI).  Solutions for each DoI are
either used locally to subtract from the data
(section~\ref{Sec:Peel}), or interpolated for other sources as well.

\begin{figure*}[ht!]
\centering
\includegraphics[width=9cm]{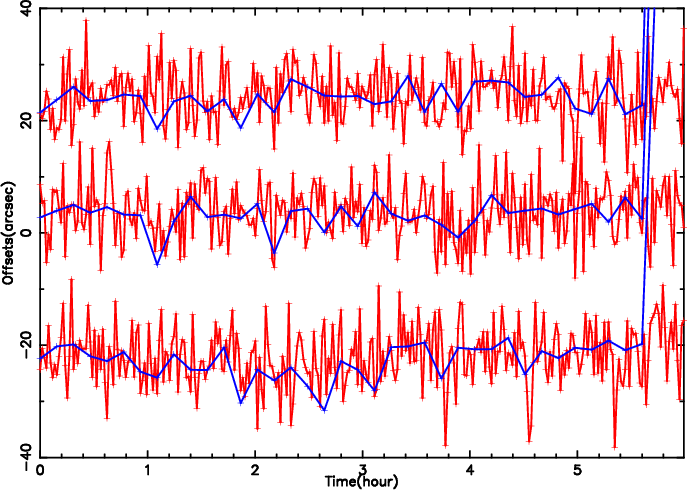}
\caption{\small Figure showing simulations with typical pointing
  errors for VLA antennas as a function time (red curves drawn with
  lines and symbols).  The solutions for antenna pointing errors
  derived using the Pointing SelfCal algorithm are the over-plotted
  curves (blue).}
\label{Fig:PointingSelfCal}
\end{figure*}

\subsection{Pointing SelfCal}
Another approach, fundamentally different from Peeling, is to develop
physical models for the various DD effects and solve for the
parametric model using Equation~\ref{ME}.  Projection methods to
correct for known DD effects (sections~\ref{Sec:WProj} and
\ref{Sec:AProj}) can be easily used to implement solvers which are
also computationally efficient.  This approach fundamentally mitigates
the problem of the proliferation of DoFs inherent in the peeling
approach.  The Pointing SelfCal algorithms \citep{POINTING_SELFCAL} is
an example of use of this approach to solve for antenna pointing
errors.  To correct for the pointing errors, the solved pointing
errors are included as part of the model for the antenna aperture
illumination and used in A-Projection algorithm during imaging.
Fig.~\ref{Fig:PointingSelfCal} shows results from tests using
simulated data.  Further work on this using real data is currently in
progress.


\acknowledgements 

I wish to thank T.J. Cornwell, K. Golap, R. Nityananda and U. Rau for
their useful comments and discussions over many years.

\bibliographystyle{asp}
\bibliography{lfru_sanb}

\end{document}